\documentclass[nofootinbib,notitlepage,superscriptaddress,10pt,aps,pra,twocolumn]{revtex4-1}

\usepackage[utf8x]{inputenc}
\usepackage{amsmath}
\usepackage{amssymb}
\usepackage{bbm}
\usepackage{graphicx}

\begin{document}


$~~~~~~~~~~~~~~~~~~~~~~~~~~~~~~~~~~~~$ {\small $preprint~version~of~Physical~Review~Letters~111~(2013)~101301$}

\title{$~~$\\Challenge to Macroscopic Probes of Quantum Spacetime\\
 Based on Noncommutative Geometry}

\author{Giovanni AMELINO-CAMELIA}
\affiliation{Dipartimento di Fisica, Universit\`a di Roma ``La Sapienza", P.le A. Moro 2, 00185 Roma, EU}
\affiliation{INFN, Sez.~Roma1, P.le A. Moro 2, 00185 Roma, EU}

\begin{abstract}
Over the last decade a growing number of quantum-gravity researchers
has been looking for opportunities for the first ever experimental evidence of a Planck-length
quantum property of spacetime. These studies are usually based on
the analysis of some candidate indirect implications of spacetime
quantization, such as a possible curvature of momentum space.
Some recent proposals have raised hope that we might also gain direct
 experimental access to quantum properties
of spacetime, by finding evidence of limitations to the measurability
of the center-of-mass coordinates of some macroscopic bodies.
However I here observe that the arguments that originally led to
speculating about spacetime quantization do not apply to the localization
of the center of mass of a macroscopic body.
And I also analyze some popular formalizations of the notion of quantum spacetime,
finding that when the quantization of spacetime is Planckian for the constituent
particles then for the center of mass of a composite macroscopic body
the quantization of spacetime is much weaker than Planckian.
These results suggest that the center-of-mass observables of macroscopic bodies should not provide
good opportunities for uncovering quantum properties of spacetime. And they also raise
some conceptual challenges for theories of mechanics in quantum spacetime,
in which for example free protons and free atoms
 should feel the effects of spacetime quantization differently.
\end{abstract}

\maketitle

\section{Introduction and Motivation}
Traditionally the quantum-gravity problem was studied as a mere technical exercise,
assuming that it might be impossible to find
experimental evidence of the minute effects produced by
the characteristic length scale of quantum gravity,
expected to be of the order of the Planck length $\ell_P \simeq 10^{-35}m$.
This changed over the last decade as a result of a growing number of studies
(see, {\it e.g.},
Refs.~\cite{grbgac,gampul,schaefer,gacgwi,mexweave,jaconature,ngPRL,piranNeutriNat,fermiNATURE,tamburini,prl})
showing that evidence of Planck-length
quantum properties of spacetime might be within our experimental reach if we
exploit some candidate {\underline{indirect}} manifestations of spacetime quantization.
An intuitive example of candidate indirect manifestations of spacetime quantization
is found in results showing that certain ways to introduce the Planck length
as scale of spacetime quantization admit a dual picture in which the Planck length
also plays the role of scale of curvature of momentum space, with implications
for relativistic kinematics (see, {\it e.g.}, Refs.~\cite{prl,majidCURVATURE}).

It would of course be important to also find opportunities
for observing Planck-length spacetime quantization {\underline{directly}}.
And according to the studies recently reported in 
Refs.~\cite{pikovski,bekenstein}\footnote{As this Letter was being finalized
 I became also aware of the proposal put forward  in
 Ref.~\cite{auriga}, which is in part analogous to the proposals
 in Refs.~\cite{pikovski,bekenstein}, from the viewpoint here adopted.
 Ref.~\cite{auriga}
 seeks evidence of spacetime quantization by exploiting the sub-millikelvin cooling
of the normal modes of the gravitational wave
detector AURIGA,
a 3-meter long aluminum bar
weighing 2.3 tons.}
 this might be possible,
at least in the sense that we can achieve Planckian accuracy
in measurements pertaining the
center-of-mass coordinates of some macroscopic bodies.
The study reported by Pikovski {\it et al} in Ref.~\cite{pikovski}
focuses on the center-of-mass motion
of a mechanical oscillator,
while the study reported  by Bekenstein in Ref.~\cite{bekenstein}
focuses on the center-of-mass motion
of a macroscopic dielectric block  traversed by a single optical photon.

In attempting to assess the likelihood of success
of these proposals I noticed that they involve small momentum transfer from a low-energy
photon to a macroscopic body, the body being describable fully within
the ``nonrelativistic limit" (small velocities, where Galilean
relativity holds).
And I find that the arguments that inspired quantum-gravity
research on Planck-length spacetime quantization do not apply to such interactions.
The current consensus among theorists (see, {\it e.g.}, the reviews in
Refs.~\cite{garayminlen,sabineminlen})
is that spacetime quantization
is needed because any attempt to localize a particle
with Planckian accuracy requires concentrating energy of order
the inverse of the Planck length within a Planck-length-size region, and in such
situations our present understanding of gravitational phenomena suggests
that a black hole should form, rendering the localization procedure meaningless.
The procedures proposed in Refs.~\cite{pikovski,bekenstein} for  Planck-length
accuracy in the control of the center-of-mass position of a macroscopic body
evidently do not involve any particularly high concentration of energy
in small regions.

The hope that the center of mass of a macroscopic body might be subject to the
same Planck-length quantum properties of spacetime expected for fundamental particles
is therefore evidently based on an implicit
inductive argument: the necessity of Planck-length spacetime quantization arises exclusively
in arguments involving
fundamental particles, but once that is accommodated in the theory perhaps by some
(unproven and unknown)
consistency criterion the Planck-length quantum properties of spacetime would also affect the center of mass of
a macroscopic body.
To my knowledge this huge extrapolation is not confirmed by any known results
of quantum-spacetime research. On the contrary I here provide a simple argument
suggesting that this extrapolation is incorrect. I consider a few of the most popular models
being studied in the quantum-spacetime literature,
and I probe conceptually the issue here at stake by using
a simplified characterization of the center of mass of a body composed of $N$ constituent particles.
I take as center-of-mass coordinates the observables $X,Y,Z$, with
\begin{equation}
X = \frac{1}{N}\sum\limits^{N}_{n=1} x_n \,\, , \,\,\, Y = \frac{1}{N}\sum\limits^{N}_{n=1} y_n
 \,\, , \,\,\, Z = \frac{1}{N}\sum\limits^{N}_{n=1} z_n
 \label{xcom}
 \end{equation}
(where of course $x_n,y_n,z_n$ are the coordinates of the n-th composing particle), and I take
 as center-of-mass momentum the observables $P_x,P_y,P_z$, with
\begin{equation}
P_x = \sum\limits^{N}_{n=1} p_{x,n}\,\, , \,\,\, P_y = \sum\limits^{N}_{n=1} p_{y,n}
\,\, , \,\,\, P_z = \sum\limits^{N}_{n=1} p_{z,n}
\label{pcom}
\end{equation}
(where of course $p_{x,n},p_{y,n},p_{z,n}$ are the momentum components of the n-th composing particle).

This simplified description of a macroscopic body is sufficient for my purposes
 since the relevant
phenomenological opportunities are for the center of mass of macroscopic bodies in the nonrelativistic regime and my
main objective is to provide a counter-example to the conjecture
that Planck-length quantum properties of spacetime
apply in undifferentiated way to fundamental particles and to the center of mass of macroscopic
bodies.
I shall show that the conjecture is false by showing that it does not apply to
macroscopic bodies whose center-of-mass motion is characterized
by (\ref{xcom})-(\ref{pcom}).
And (\ref{xcom})-(\ref{pcom}) is appropriate for
macroscopic bodies whose constituents  all have the same mass
and whose center-of-mass degrees of freedom  decouple
from the other degrees of freedom.

\section{Results for classical spacetime and Lie-algebra quantum spacetime}
Let me first recall the mechanism through which the
description (\ref{xcom})-(\ref{pcom}) gives satisfactory results within
ordinary quantum mechanics, in classical spacetime, where the only non-trivial commutator
is Heisenberg's
$$[x,p_x]=i \hbar$$
(focusing for simplicity on the $x$-direction).

Evidently the Heisenberg commutator also applies to a body's center of mass
described by (\ref{xcom})-(\ref{pcom}):
\begin{eqnarray}
[X,P_x] &=& \left[\frac{1}{N}\sum\limits^{N}_{n=1} x_n \, , \sum\limits^{N}_{m=1} p_{x,m} \right] \\
 &=& \frac{1}{N} \sum\limits^{N}_{n=1} \sum\limits^{N}_{m=1} \delta_{n,m} i \hbar
 =  \frac{1}{N} \sum\limits^{N}_{n=1}  i \hbar = i \hbar \nonumber
\end{eqnarray}

My next application is already non-trivial and novel, but nonetheless provides further
elements in support of the usefulness of the conceptual probe I am using,
centered on (\ref{xcom})-(\ref{pcom}).
For this I consider a class of quantum-spacetime pictures involving noncommutativity of coordinates of Lie-algebra type~\cite{majrue,lukieANNALS,bala}
$$[r^\alpha,r^\beta]=i \ell \theta^{\alpha \beta}_\gamma r^\gamma $$
with\footnote{I focus on spatial noncommutativity, which suffices
for establishing the issue for macroscopic bodies which is here of interest.}
$r^1=x$, $r^2=y$, $r^3=z$.\\
This type of noncommutativity of coordinates
is here particularly significant since
it is the only case where the literature does provide a suggestion that macroscopic bodies
might be affected by Planck-length features differently from their constituent particles.
These are arguments focusing on the description of macroscopic bodies when momentum space
is curved or anyway affected by nonlinearities (see Ref.~\cite{soccerball} and references therein).
Lie-algebra spacetimes are known to be dual to momentum spaces with curved geometry~\cite{prl,majidCURVATURE}
and one of the implications is that the laws of conservation of momentum for fundamental particles
are Planck-length deformed. Applying the relevant deformed
conservation laws to the constituents of a macroscopic body can give a net result for collisions
 such that momentum conservation for macroscopic-body total momentum
is affected by weaker corrections than momentum conservation for the particle constituents.
Specifically, Ref.~\cite{soccerball} focused on a situation such that before and
after the momentum exchange the bodies are composed of particles in exactly rigid motion  and found that the curvature
of momentum space was felt by the macroscopic body not as set by the Planck length but rather as set by the Planck length
divided by the number $N$ of particle constituents.

Even though they applied only to rather special contexts
(exact rigid motion is an assumption stronger than
the ones required by my Eqs.(\ref{xcom})-(\ref{pcom}))
and they concerned momentum-space nonlinearities rather than spacetime fuzziness,
these previous arguments could already hint at the possibility that
in Lie-algebra spacetimes the effective Planck length should be rescaled
for macroscopic bodies.
My simple ``conceptual probe" produces for the noncommutativity of coordinates results which are indeed
consistent with the intuition emerging from those previous studies on the momentum-space side.
To see this let me consider
the case of a commutator of type
$$[x,y]=i \ell r^\alpha $$
with $\alpha$ taking any value among $1,2,3$ (so that essentially I consider at once cases
of the type $[x,y]=i \ell x $ and of the type $[x,y]=i \ell z $).

Applying $[x,y]=i \ell r^\alpha $ to the constituent particles of a macroscopic
body one then finds for  the center-of-mass coordinates described in  (\ref{xcom}) the result
\begin{eqnarray}
[X,Y] & = & \left[\frac{1}{N}\sum\limits^{N}_{n=1} x_n \, , \frac{1}{N}\sum\limits^{N}_{m=1} y_m \right] \label{liealgebrascaling}\\
 & = & \frac{1}{N^2} \sum\limits^{N}_{n=1} \sum\limits^{N}_{m=1} \delta_{n,m} i \ell r^\alpha_n
 =  \frac{1}{N^2} \sum\limits^{N}_{n=1}  i \ell r^\alpha_n = i  \frac{ \ell}{N}  R^\alpha
\nonumber
\end{eqnarray}
where of course $R^\alpha \equiv N^{-1} \sum_{n=1}^N r^\alpha_n$.\\
Evidently (\ref{liealgebrascaling}) shows that the effects of Lie-algebra  coordinate noncommutativity
for the center of mass of macroscopic bodies are scaled down by a factor of $1/N$. While this could be expected intuitively
on the basis of the dual momentum-space picture described in Ref.~\cite{soccerball}, it is noteworthy that my approach
provides a consistent picture of the quantum-spacetime aspects.

\section{Results for other quantum-spacetime pictures}
I shall now show that my perspective on center-of-mass degrees of freedom of  macroscopic bodies
has applicability that goes beyond the specific context of Lie-algebra spacetime noncommutativity.
My next example is ``Moyal noncommutativity", with coordinate-independent
commutators, such as
\begin{equation}
[x,y]=i \ell_M^2
\label{moyal}
\end{equation}
This is perhaps the most studied candidate scenario for the quantization of spacetime
(see, {\it e.g.}, Refs.~\cite{doplicherFR,szabo} and references therein), and
 to my knowledge there is no result in
the literature\footnote{Though there is no trace of it in the literature,
credit for Eq.(\ref{moyalMAIN}) should go to 
Volovik.
In private conversations motivating his approach to quantum gravity~\cite{volovik},
Volovik argued, as early as 2003, that Eq.(\ref{moyalMAIN}) would cast a shadow on
Moyal noncommutativity. 
I became convinced of the significance of
Eq.(\ref{moyalMAIN}) at the end of  a path that took me 
first to results
on the dual momentum-space picture  of some Lie-algebra
spacetimes~\cite{soccerball}, and then to Eq.(4)
for Lie-algebra spacetimes, whose consistency with
the findings of Ref.~\cite{soccerball} was a key
source of encouragement for going forward. 
I look at Eq.(6) 
(and Eq.(4)) as
a challenge which could also turn (see later) into an exciting opportunity.}
 anticipating that macroscopic bodies should
be affected by Moyal noncommutativity differently from
their constituents. 
The applicability
of my thesis to Moyal noncommutativity is easily
checked by using (\ref{xcom}) for center-of-mass coordinates with
the constituents governed by noncommutativity (\ref{moyal}):

\begin{eqnarray}
&&[X,Y] = \left[\frac{1}{N}\sum\limits^{N}_{n=1} x_n \, , \frac{1}{N}\sum\limits^{N}_{m=1} y_m \right] \label{moyalMAIN}\\
 && \,\,\,\,\,\, = \frac{1}{N^2} \sum\limits^{N}_{n=1} \sum\limits^{N}_{m=1} \delta_{n,m} i \ell_M^2
 =  \frac{1}{N^2} \sum\limits^{N}_{n=1}  i \ell_M^2
 = i \left(\frac{\ell_M}{\sqrt{N}}\right)^2 \nonumber
\end{eqnarray}
Therefore also for the Moyal case
the noncommutativity of center-of-mass coordinates
should be weaker than the noncommutativity
of the coordinates of the constituents. Specifically the Moyal
noncommutativity length scale $\ell_M$ gets reduced by a factor
of $1/\sqrt{N}$.

Another much studied class of quantum-spacetime pictures that I should consider is the
one that does not invoke noncommutativity of coordinates, but is instead
centered on modifications of the Heisenberg commutator of the general type~\cite{kempf,vagenas}
\begin{equation}
[x,p]=i \hbar (1 - \lambda' p + \lambda^2 p^2)
\label{achimlike}
\end{equation}
Even with commuting coordinates these modifications of the Heisenberg commutator
produce spacetime quantization. The key role for this is played
 by the parameter $\lambda^2$ of the quadratic term. The standard Heisenberg commutator still allows
localizing a particle sharply at a point ($\delta x \rightarrow 0$)  if $\delta p \rightarrow \infty$,
 {\it i.e.} if all information on the conjugate momentum is given up.
But for  $\lambda^2 \neq 0$ the Eq.~(\ref{achimlike}) produces a see-saw formula~\cite{kempf,vagenas}
such that $\delta x$ receives a novel contribution proportional $\delta p$
in addition to the standard Heisenberg term going like $1/\delta p$, in such a way
that
the coordinate $x$ cannot ever be measured sharply, as required
for a quantum-spacetime picture.\\
Of some interest for my thesis is also the perspective given in
Ref.~\cite{vagenas}, advocating the specific choice of $\lambda' = \lambda$
in (\ref{achimlike}), partly because of its consistency (in the sense of Jacobi identities) with commutativity
of coordinates among themselves and of momenta among themselves.

Keeping these facts in mind it is then interesting to look at the properties of a center of mass described by (\ref{xcom})-(\ref{pcom})
when the constituents are governed by (\ref{achimlike}):
\begin{eqnarray}
[X,P_x] &=& \left[\frac{1}{N}\sum\limits^{N}_{n=1} x_n \, , \sum\limits^{N}_{m=1} p_{x,m} \right] \label{defheisMAIN}\\
 &=& \frac{1}{N} \sum\limits^{N}_{n=1} \sum\limits^{N}_{m=1} \delta_{n,m} i \hbar (1 - \lambda' p_{x,m} + \lambda^2 p_{x,m}^2)\nonumber\\
 &=& i \hbar \left[1 -  \frac{\lambda'}{N} P_x
 + \frac{\lambda^2}{N^2}P_x^2
 + \frac{\lambda^2}{N} \sum\limits^{N}_{n=1} \left(p_{x,n}^2 - \frac{P_x^2}{N^2} \right) \right] \nonumber\\
&\simeq &  i \hbar \left( 1-  \frac{\lambda'}{N} P_x  +   \frac{\lambda^2}{N^2} P_x^2 \right) \nonumber
\end{eqnarray}
where for the last approximate equality I restricted my attention to macroscopic bodies in (quasi-)rigid motion, as those of interest for the mentioned
experimental proposals put forward in Refs.~\cite{pikovski,bekenstein},
so that one can expect for every $n$
that $p_{x,n}  \simeq P_x/N$. Evidently, at least in this rigid-motion limit,
also for quantum spacetimes characterizable in terms
of Eq.(\ref{achimlike})
I am finding that the center of mass of  a macroscopic body should be affected more weakly than its constituents by spacetime quantization.
Notably my argument suggests that in the rigid-motion limit the
length scales in Eq.(\ref{achimlike}), both $\lambda'$ and $\lambda$, get
scaled down by $1/N$. This appears to ensure in particular that the prescription  $\lambda' = \lambda$ advocated in
Ref.~\cite{vagenas} could apply both to fundamental particles and to
the center of mass of a macroscopic body in rigid motion (but in the macroscopic case both  $\lambda'$
and $\lambda$ are reduced by $1/N$).

It is also important to notice that for bodies whose motion is not well approximated as rigid
there could be significant changes to the scaling with $N$ of the quantum-spacetime effects,
because the last equality in
Eq.(\ref{defheisMAIN}) would be inapplicable. This evidently does not introduce a limitation for
the thesis I am here presenting: even in cases where it can be expected that the quantum-spacetime
effects do not scale exactly by the number $N$ of constituents
it would of course still be wrong to fall back on the naive assumption assigning the same quantum-spacetime
properties to center-of-mass degrees of freedom of a macroscopic body and to the degrees of freedom
of its constituent particles. A dedicated technical analysis will be needed in each case, and interestingly in
some cases such a careful analysis could be motivated even by phenomenological
prospects.
Particularly significant from this perspective could be studies of macroscopic
bodies at ultra-high temperature\footnote{I am grateful to two anonymous referees for pointing my attention
to the possible role of the term involving $\sum (p_{x,n}^2 - P_x^2/N^2)$ in the description
of ultrahot bodies.}.
The
term  $N^{-1} \sum\limits^{N}_{n=1} (p_{x,n}^2 - P_x^2/N^2)$,
neglected for the last equality in
Eq.(\ref{defheisMAIN}), might indeed be large for ultrahot bodies,
for which the deviations $p_{x,n}^2 - P_x^2/N^2$ are typically large. This is not the case of
the macroscopic bodies considered
in the phenomenological proposals
of Refs.~\cite{pikovski,bekenstein}, but could inspire some new phenomenological proposals, as I shall stress
in parts of the next section.

\section{Implications and outlook}
The analysis I here reported should put to rest any further temptations of relying on the unquestioned assumption
that the center-of-mass degrees of freedom of a macroscopic body be affected by quantum-spacetime effects
just as much as the microscopic constituents of the body.
I have provided counter-examples to that assumption which, because of the nature
of my conceptual probe
centered on (\ref{xcom})-(\ref{pcom}), are robust at least for the center of mass of
bodies in quasi-rigid motion (like a solid a low temperatures).
Let me also stress that it does not take a particularly macroscopic system for my concerns to be
applicable. Think of just bound systems of two identical particles, with coordinate
vectors ${\vec{r}}_1$ and ${\vec{r}}_2$ and with bounding potential $V(|{\vec{r}}_1 - {\vec{r}}_2|)$ affecting
only the relative motion:
for such systems (\ref{xcom}) and (\ref{pcom}) are correct, with $N=2$.

I feel that the pattern I here exposed for the description of center-of-mass degrees of freedom
of macroscopic bodies in quantum spacetime could actually teach us more than
the inadequacy of previous assumptions. I stumbled upon structures which are to a large extent
similar in the study of ``Lie-algebra noncommutativity", ``Moyal noncommutativity"
and ``Deformed-Heisenberg quantum spacetimes". Readers familiar with the related literature will appreciate
that these three classes of models reflect widely different intuitions and formalizations of the quantum-spacetime
notion, and it is therefore surprising that such a consistent pattern did arise.
Among the most studied quantum-spacetime pictures the most noticeable omission in my list
is Loop Quantum Gravity~\cite{rovelliLRR2008}, and
it would of course be interesting to generalize my argument to
Loop Quantum Gravity. 

While I am proposing that simplistic assumptions about the properties of macroscopic bodies in a quantum
spacetime must be abandoned for good, I believe it would be incorrect to give up on the idea of discovering quantum-spacetime effects
through observations of macroscopic bodies. After all (if only the development of observational techniques had had a different history)
 quantum mechanics itself could have been discovered by studying white dwarfs, rather than through observations
 at atomic and subatomic scales. There might be out there an opportunity for uncovering a manifestation of quantum spacetime
 through studies of some specific macroscopic bodies. But in order for us to capitalize from such opportunities it will be
 necessary to move much beyond simple-minded assumptions about general properties of center-of-mass degrees of freedom.
Macroscopic bodies have a huge variety of properties, and only some special ones among them under some suitable
 special conditions (and for observables not necessarily linked to the center-of-mass degrees of freedom)
 could manifest quantum-spacetime properties tangibly.

 One could try with macroscopic bodies for which the center-of-mass
 degrees of freedom do not fully decouple from the internal degrees of freedom. In such cases the arguments I here reported would be inapplicable,
 but of course this does not mean that some naive guess work is then allowed. One should handle the tough challenge of modeling such bodies
 and figure out under which conditions the Planck-scale effects could be tangible. And it will be necessary to achieve
 rigorous quantifications of the implications
 for a given macroscopic body of interest: in phenomenology also negative results are important since they allow to set limits on
 the parameters of candidate new theories, but that is only possible if the quantification of predicted effects is rigorously derived
from the defining parameters of the theory.

Similar considerations can be inspired by the contributions of
type $p_{x,n}^2 - P_x^2/N^2$ neglected for the last equality in Eq.(\ref{defheisMAIN}) under the assumption
of quasi-rigid motion. My Eq.(\ref{defheisMAIN}) also shows that for ``deformed-Heisenberg noncommutativity" one could have an amplification
of the quantum-spacetime effects when
the body is not quasi-rigid and the context is such that terms of type $p_{x,n}^2 - P_x^2/N^2$ are large. This in particular should be expected
 for bodies at particularly high temperatures. But notice that the properties of the center of mass of bodies in such extreme regimes
would still be different from the ones of the constituents. For appropriately large departures from
quasi-rigid motion in deformed-Heisenberg quantum spacetimes the Planck-scale properties of the center of mass of a
macroscopic body could actually be stronger than those of the constituents.

It is also possible that for some models of quantum spacetime the starting points of my analysis,
constituted by (\ref{xcom}) and (\ref{pcom}), are inapplicable even when the center-of-mass degrees of freedom cleanly decouple from
internal degrees of freedom. For one of the cases here considered, the one of Lie-algebra noncommutative spacetime, this is already
established in the literature, though it does not affect my analysis. Indeed in Lie-algebra spacetimes the
law of composition of momenta is expected to be deformed but the law of composition of spacetime coordinates is undeformed, as first shown in Ref.~\cite{majrue}.
Interestingly the derivation of my main result for Lie-algebra spacetimes, Eq.~(\ref{liealgebrascaling}),
requires exclusively (\ref{xcom}), so it is not affected by this issue. The requirement (\ref{pcom}) is crucial for my
main result
 concerning ``deformed-Heisenberg noncommutativity", Eq.(\ref{defheisMAIN}), but the available literature on those quantum spacetimes does not
 advocate any deformation of composition laws
 (see, {\it e.g.}, Refs.~\cite{kempf,vagenas}).
 Similarly the available literature on ``Moyal noncommutativity", for which my main result is Eq.~(\ref{moyalMAIN}),
  does not advocate~\cite{doplicherFR,szabo} any modification of (\ref{xcom}) and (\ref{pcom}).
 So the analysis I here reported is not challenged by any available results on composition laws in quantum spacetimes. However,
  this issue must be monitored since the understanding of known quantum-spacetime models is still in progress.
   Moreover, new models might at some point be proposed with deformed composition laws such that
 my argument would then not be applicable to them.

While my main focus here was on phenomenological prospects, in closing I should also emphasize
some severe technical challenges that, according to my analysis, must be faced in theory work on the quantum-spacetime idea.
A first challenge comes from the fact that my analysis shows that macroscopic bodies
have quantum-spacetime properties different from those of their constituents, but it gives
no indication of which constituents are those ``fundamental enough" to be affected by the full strength
of Planck-scale effects. Think for example of molecules: my analysis suggests that molecules are affected more weakly
by quantum-spacetime effects than the atoms within them, but Planck-length magnitude of quantum-spacetime
effects should be assumed for atoms or for protons and neutrons within the nuclei of atoms? or for quarks?\\
And a second challenge would need to be  faced
even assuming this first challenge is eventually addressed in a given quantum-spacetime
picture, so that actually the picture predicts the magnitude of quantum-spacetime effects for,
say, protons and also predicts how much weaker than for protons the effects are for, say, Cs atoms.
We would clearly need a completely new type of theory of mechanics,
in which the spacetime properties of different particles are different. We should renounce to one
of the key aspects of simplicity that survived previous evolutions of our formulation of the laws of physics:
the general-relativistic description of spacetime, just like the special-relativistic one and the Newtonian one,
is indeed such that the implications of spacetime for particle properties are independent of compositeness,
and are therefore the same for protons and large atoms.

\bigskip

\bigskip

\bigskip

I gratefully acknowledge valuable conversations with L.~Freidel, A.~Grillo, 
 J.~Kowalski-Glikman,
  I.~Pikovski  and L.~Smolin.
This work was
supported in part by a grant from the John Templeton Foundation.

\end{document}